\documentclass[conference]{IEEEtran}
\usepackage{cite}

\ifCLASSINFOpdf
  \usepackage[pdftex]{graphicx}
\else
\fi
\usepackage[cmex10]{amsmath}
\usepackage[tight,footnotesize]{subfigure}

\usepackage{stfloats}

\usepackage{url}

\begin{document}
%
\title{DeepJ: Style-Specific Music Generation}

\author{\IEEEauthorblockN{Huanru Henry Mao}
\IEEEauthorblockA{Computer Science and Engineering\\
University of California, San Diego\\
9500 Gilman Dr, La Jolla, CA 92093 \\
Email: hhmao@ucsd.edu}
\and
\IEEEauthorblockN{Taylor Shin}
\IEEEauthorblockA{Computer Science and Engineering\\
University of California, San Diego\\
9500 Gilman Dr, La Jolla, CA 92093 \\
Email: t2shin@ucsd.edu}
\and
\IEEEauthorblockN{Garrison W. Cottrell}
\IEEEauthorblockA{Computer Science and Engineering\\
University of California, San Diego\\
9500 Gilman Dr, La Jolla, CA 92093 \\
Email: gary@ucsd.edu}}


%


\maketitle

\begin{abstract}
Recent advances in deep neural networks have enabled algorithms to compose music that is comparable to music composed by humans. However, few algorithms allow the user to generate music with tunable parameters. The ability to tune properties of generated music will yield more practical benefits for aiding artists, filmmakers, and composers in their creative tasks. In this paper, we introduce DeepJ - an end-to-end generative model that is capable of composing music conditioned on a specific mixture of composer styles. Our innovations include methods to learn musical style and music dynamics. We use our model to demonstrate a simple technique for controlling the style of generated music as a proof of concept. Evaluation of our model using human raters shows that we have improved over the Biaxial LSTM approach.
\end{abstract}


%
\IEEEpeerreviewmaketitle

\section{Introduction}
Music composition is a challenging craft that has been a way for artists to express themselves ever since the dawn of civilization. Designing algorithms that produce human-level art and music, in the field of artificial intelligence, is a difficult yet rewarding challenge. Recently, advances in neural networks have helped us transition from writing music composition rules to developing probabilistic models that learn empirically-driven rules from a vast collection of existing music.

Neural network music generation algorithms have been limited to particular styles of music, such as jazz, Bach chorales, and pop music~\cite{schmidhuber, deep_bach,song_from_pi}. These music generation methods have created interesting compositions, but their specialization to a particular style of music  may be undesirable for practical applications. Genre-agnostic methods of composing music, such as the Biaxial LSTM \cite{biaxial} are able to train using a variety of musical styles, but are unable to be style-consistent in their outputs, pivoting to different styles within one composition.

In this paper, we introduce DeepJ\footnote{Our code is available on \url{https://github.com/calclavia/DeepJ/tree/icsc} for reference.} (a reference to disc jockeys or DJs), a deep learning model capable of composing polyphonic music conditioned on a specific or a mixture of multiple composer styles. Our main contributions in this work are the incorporation of enforcing musical style in the model's output and learning music dynamics. In the context of this paper, we refer to style as either a music genre or a particular composer's style. Due to data limitations, our paper focuses on composers from different classical eras, but we believe the techniques presented here generalizes to other types of music. In music theory, {\it dynamics} is the volume of each note, here ranging from \textit{pianissimo} to \textit{fortissimo}.

We believe a model that is capable of generating music in various styles will yield practical benefits for filmmakers and music composers who need to customize generated music for their creative tasks. For example, a filmmaker may wish to match a video with music that is of a particular style to convey a desired emotion. Our method of incorporating style serves as a proof of concept of this idea. Our technique could be extended to other tunable parameters in generated music such as ``mood'' or ``emotion.''

\section{Background}
\subsection{Previous Work}
Monophonic music generation focuses on the task of generating a melody - a single tune without harmony. Early work such as \textit{CONCERT} \cite{mozer} attempted to generate melody by estimating the probability of playing the next note as a function of previous notes. As Mozer pointed out, the advantage of recurrent networks is that they are not restricted to Markovian predictions, but the memory capacity of simple backpropagation through time is practically limited. More recent work \cite{schmidhuber} has improved on this by using Long-Short Term Memory (LSTM) units \cite{lstm} that can make predictions based on distal events. In summary, these methods attempt to learn melody composition by modeling the note to play next probabilistically, conditioned on previously generated notes:
\[
P(Y_{n+1}=y_{n+1} | Y_n=y_n, Y_{n-1}=y_{n-1}, Y_{n-2}=y_{n-2}, ... )
\]

Polyphonic music generation is more complex than monophonic music generation. In each time step, the model needs to predict the probability of any combination of notes to be played at the next time step. Early work in polyphonic composition involving neural networks attempted to model sequences using a combination of RNNs and restricted Boltzmann machines (RBM) \cite{modeling_temporal_dependencies}. This work demonstrated excellent results by using an RBM to generate a distribution over the notes in each time step, conditioned on the previous time step. More recent architectures such as the Biaxial LSTM describe a type of deep learning that is able to model a joint probability distribution of notes with transposition invariance \cite{biaxial}.

\subsection{Biaxial LSTM}
\subsubsection{Data Representation}
The Biaxial LSTM architecture uses a piano roll representation of notes, which is a dense representation of MIDI (a digital score format) music. A piano roll represents the notes played at each time step as a binary vector, where a 1 represents the note corresponding to its index is being played and a 0 represents the note is not being played. A piece of music is a \(N \times T\) binary matrix where \(N\) is the number of playable notes and \(T\) is the number of time steps. The following is an example of representing two notes held for two time steps followed by two time steps of silence in a representation that captures \(N = 4\) notes.

\[
t_{play} = 
\begin{bmatrix}
    0 & 0 & 0 & 0 \\
    1 & 1 & 0 & 0 \\
    0 & 0 & 0 & 0 \\
    1 & 1 & 0 & 0
\end{bmatrix}
\]

Only representing note play is insufficient in capturing all MIDI actions - there is a difference in holding a note versus replaying a note. Note replay is defined as the event when a note is re-attacked immediately after the note ends, with no time steps in between successive plays. This is defined by a replay matrix \(t_{replay}\) similar to \(t_{play}\). Note play and replay jointly defines the note representation.

\subsubsection{Architecture}
The Biaxial LSTM architecture \cite{biaxial} models polyphonic music by modeling each note within each time step as a probability conditioned on all previous time steps and all notes within the current time step that have already been generated. The order of generation is from the first time step \(t = 1\) to the last time step \(t = T\) and from the lowest note \(n = 1\) to the highest note \(n = N\), where \(T\) is the length of desired generation and \(N\) is the number of possible notes. We denote the binary random variable \(Y = 1\) when the note is played and \(Y = 0\) otherwise. At time step \(t\) the model learns the following conditional probability for the \(n\)-th note:
\[
P(Y_{t,n}|Y_{t,n-1},Y_{t,n-2},...,Y_{t-1,N},Y_{t-1,N-1},...,Y_{1,2},Y_{1,1})
\]

From the expression above, we can see that the model must condition the probability of playing a note along two axes: time and note. The model performs this biaxial probability conditioning by having LSTMs that take inputs along the different axes respectively. Hence, the Biaxial LSTM consists of two primary modules: the time-axis module and note-axis module.

\begin{figure}
  \centering
  \includegraphics[width=0.5\textwidth]{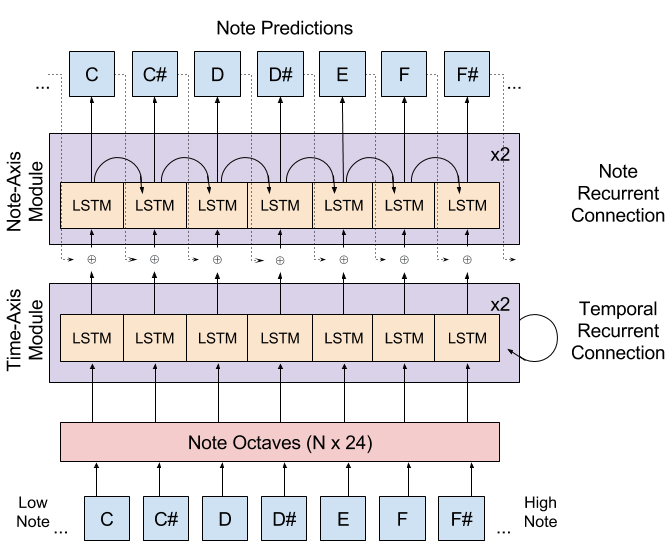}
  \caption{Biaxial LSTM Architecture - The \(\oplus\) symbol represents concatenation. The x2 represents the module being stacked twice. }
  \label{fig:biaxial}
\end{figure}

The input to the model are notes in the previous time step. The "note octaves" (Figure \ref{fig:biaxial}) layer transforms each note into a tensor of the note and its surrounding octave of notes, augmenting the note with spatial context. Every note feature is fed into an LSTM with shared weights across notes in the time-axis. The time-axis module takes note octaves and recurrent states from the previous time step and outputs higher level note features for each note.

The primary advantage of shared LSTM weights in the time-axis module is transposition invariance. In music, harmonies and melodies are determined more by their relative pitches rather than absolute pitches - music transposed from one key to another is still considered the same music. This is analogous to translation invariance in image classification.
%
%

The time-axis section is inspired by a convolution neural network. It consists of two layer stacked LSTM units recurrent in time connected to each note octave (12 pitches above and below every note) in the input in a fashion similar to a convolution kernel. The weights of each LSTM unit are shared across each note, forcing the time-axis section to learn note invariant features. This allows the module to generalize to different transpositions. The time axis section outputs features for each note and can be computed in parallel.

Taking in the note feature outputs from the time axis as input, the note-axis LSTM sweeps from the lowest note feature to the highest note feature to make predictions of each note conditioned on the predicted lower notes. The note-axis consists of another two layer stacked LSTM that is recurrent in note. Each note's features are first  concatenated with the lower chosen note. If the current note being considered is the lowest note, then zeros are concatenated. In other words, to determine the chosen note \(y_{i}\), the \(i\)-th note's features \(x_i\) is concatenated with the lower chosen note \(y_{i-1}\) before feeding into note-axis module \(f\).

\[
y_i = sample(f(x_i \oplus y_{i-1}))
\quad
y_1 = sample(f(x_i \oplus \boldsymbol{0}))
\]

where \(\oplus\) denotes the concatenation operator and the \(sample\) function we use samples the sigmoid probability output using a coin flip. This method of providing previous chosen notes enables the note-axis LSTM to learn probability estimations based on lower chosen notes.

The biaxial architecture also feeds contextual inputs to each LSTM in the time-axis (Figure \ref{fig:deepj}), which includes a representation of the current beat in the bar. The representation used gives the position of the time step relative to a 4/4 measure in binary format \cite{biaxial}. We found that such contextual input is necessary in aiding the model to produce outputs that are consistent in tempo. In our model we use a different representation of beat that achieves the same effect, which we describe in the following section.

\section{DeepJ}
The Biaxial architecture has created musically plausible results with measure-level structure. We build upon the architecture and introduce a simple and intuitive method for learning and enforcing musical style. We also introduce learned music dynamics in our note's representation. In the next subsections, we introduce the data representation and preprocessing step used to train the model and present our model architecture.

\subsection{Data Representation}
DeepJ uses the same note representation as the Biaxial architecture except that we augment the representation with music dynamics. We use most of the contextual inputs described in Biaxial LSTM's paper, but use an improved method of representing beat.
%
%

Dynamics in music is defined as the relative volume of a note. MIDI files contain information about how loud a note is (dynamics ranges from 0 to 127) in each note event. We keep track of the dynamics of every note in an  \(N \times T\) dynamics matrix that, for each time step, stores values of each note's dynamics scaled between 0 and 1, where 1 denotes the loudest possible volume. The following is an example of holding the same two notes from the previous example for two time steps at 0.4 volume.

\[
t_{dynamics} = 
\begin{bmatrix}
    0 & 0 & 0 & 0 \\
    0.4 & 0.4 & 0 & 0 \\
    0 & 0 & 0 & 0 \\
    0.4 & 0.4 & 0 & 0
\end{bmatrix}
\]

In our preliminary work, we also tried an alternate representation of dynamics as a categorical value with 128 bins as suggested by \textit{Wavenet} \cite{wavenet}. Instead of predicting a scalar value, our model would learn a multinomial distribution of note dynamics. We would then randomly sample dynamics during generation from this multinomial distribution. Contrary to \textit{Wavenet}'s results, our experiments concluded that the scalar representation yielded results that were more harmonious.

We collected a dataset of 23 different music composers, each of which constitute a particular artistic style. The style of the music is encoded as a one-hot representation over all artists. During generation, we mix various styles by changing the vector representation of style to include desired styles and normalize the vector to sum to one. For example, the representation of composer 1, composer 2, and the mixture of both are as follows:

\[
s_1 = [1, 0, 0, ...]\quad
s_2 = [0, 1, 0, ...]\quad
s_{mix} = [0.5, 0.5, 0, ...]
\]

We group several composers into a particular music genre. For example, if the baroque genre consists of work made by composers 1 to 4, we denote baroque as:

\[s_{baroque} = [0.25, 0.25, 0.25, 0.25, 0, 0, ...]\]

We also encode several contextual inputs to guide the model to create music with better long term structure. We feed the model the beat position of the current bar as a one-hot vector with dimension equal to $q$, the number of quantizations per bar. Note that this representation does not make any assumptions on time signature. Using $b_i$ to represent the one-hot vector with a 1 at position $i$, every bar in the music cycles through the set of all beat vectors \(b_1,b_2, ... ,b_q\). We chose $q=16$ to capture fast notes without being too computationally expensive.

\subsection{Objective Functions}
DeepJ attempts to generate music notes and dynamics. For each time step at each note, our model produces three outputs: play probability, replay probability and dynamics. We train all three outputs simultaneously. Play and replay are treated as logistic regression problems trained using binary cross entropy, as defined in Biaxial LSTM \cite{biaxial}. Dynamics is trained using mean squared error. Thus, we introduce the following loss functions:

\[L_{play} = \sum t_{play}\log (y_{play}) + (1-t_{play})\log (1-y_{play})\]

In our experiments, a naive definition of \(L_{replay}\) (\(L_{r}\) for short) and \(L_{dynamics}\) yields poor results - the dynamics tend to be near zero and replay had a low probability. We discovered that a better method is to mask out the loss for \(L_{r}\) and \(L_{dynamics}\) whenever the note is not being played, thereby setting the losses to zero. If a note is not played, replay and dynamics are never used during generation, and thus it is unnecessary to impose learning constraints on those values when they are not used.

\[
L_{r} = \sum t_{play}(t_{r}\log (y_{r}) + (1-t_{r})\log (1-y_{r}))
\]
\[L_{dynamics} = \sum t_{play}(t_{dynamics}-y_{dynamics})^2\]

\subsection{Architecture}
We base our model architecture on Biaxial LSTM's design. The primary difference between our architecture and that of Biaxial LSTM is the use of style conditioning at every layer.

Musical styles are not necessarily orthogonal to each other. For example, a classical and baroque piece likely share many characteristics. Hence, we believe representing style using a learned distributed representation is more appropriate than a one-hot representation provided by the input. We use a linear hidden layer, represented as \(W\), to linearly project the one-hot style input \(\boldsymbol{s}\) to a style embedding \(\boldsymbol{h}\).

\[
\boldsymbol{h} = W\boldsymbol{s}
\]

In our initial experiments, we directly fed this style embedding as a contextual input to the network. However, we discovered that during generation the network tends to ignore the style embedding and fails to produce music faithful to the style. Inspired by a conditioning technique from \textit{Wavenet} \cite{wavenet}, we introduce global conditioning to the Biaxial architecture.

\begin{figure}[ht]
  \centering
  \includegraphics[width=0.5\textwidth]{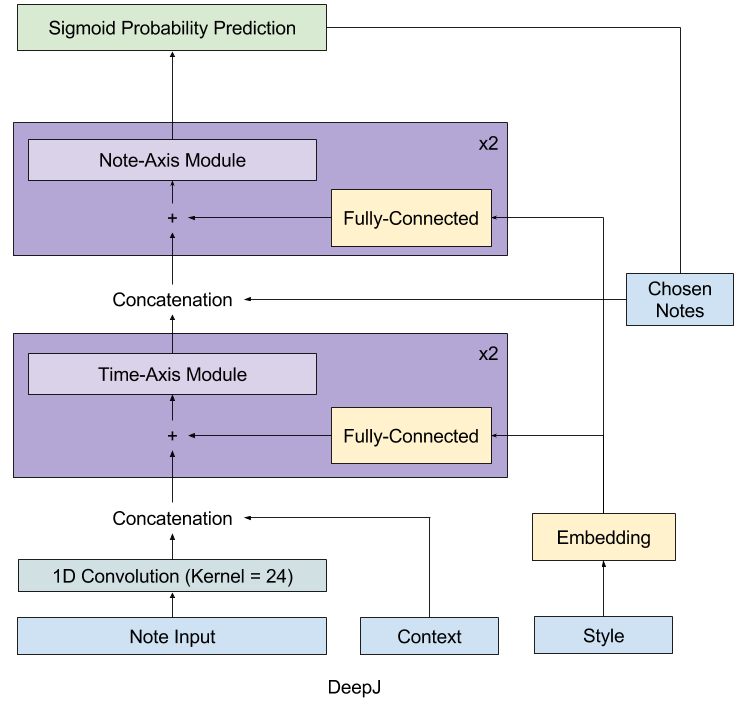}
  \caption{DeepJ Architecture}
  \label{fig:deepj}
\end{figure}

For each LSTM layer, we connect the style embedding \(\boldsymbol{h}\) to another fully-connected hidden layer with tanh activation to produce a latent non-linear representation of style \(\boldsymbol{h'}\) (yellow boxes in Figure \ref{fig:deepj}).

\[
\boldsymbol{h'_l} = tanh(W'_l\boldsymbol{h})
\]

where $l$ indexes the LSTM layer. This single latent representation influences all notes before they are fed into LSTM layers through summation. In order to make the two layers' dimensionality compatible for summation, we broadcast the latent style vector \(\boldsymbol{h'}\) across notes. In other words, for the \(i\)-th note and its input features \(\boldsymbol{x_i}\), the LSTM output \(\boldsymbol{z_i}\) for that note is defined as
 
\[
\boldsymbol{z_i} = f(\boldsymbol{x_i} + \boldsymbol{h'})
\]

where $\boldsymbol{x_i}$ is the unmodified input, "+" refers to component-wise sum, and \(f\) is the function that applies the LSTM layer to some input. This applies the same style conditioning for each note.

We also compared concatenation and summation as layer joining methods and discovered that summation provided stronger global effect on the network. In addition, we also modified the original Biaxial architecture by introducing a minor improvement. We used a 1-dimensional convolution layer to extract note features from each note's octave neighborhood. We found that this slightly improved the training loss compared to Biaxial's method of directly feeding a note's neighboring octave to the LSTM layers.

\section{Experiments}
\subsection{Training}
In our experiments, we trained DeepJ on 23 composers from three major classical periods (baroque, classical and romantic) using the same dataset\footnote{The training data can be found on \url{http://www.piano-midi.de/}.} the original Biaxial architecture was trained on. Our training set consists of MIDI music from composers ranging from Bach to Tchaikovsky. MIDI files come with a standard pitch range from 0 to 127, which we truncate to range from 36 to 84 (4 octaves) in order to reduce note input dimensionality. We also quantized our MIDI inputs to have a resolution of 16 time steps per bar.

Training was performed using stochastic gradient descent with the Nesterov Adam optimizer \cite{adam}. Specific hyperparameters used were 256 units for LSTMs in the time-axis and 128 units for the LSTMs in the note axis. We used 64 dimensions to represent our style embedding space and 64 filters for the note octave convolution layer. We used truncated back-propagation up to 8 bars back in time.

Dropout \cite{dropout} is used as a regularizer for DeepJ during training. Our model uses a 50\% dropout in all non-input layers (except the style embedding layer) and a 20\% dropout in input layers. We notice that the inclusion of dropout during training helps the model recover from mistakes it makes during generation as it becomes less dependent on the specific inputs.

\subsection{Generation}
After training, we generated samples for each genres of music. We performed generation by sampling from the model's probability distribution using a coin flip to determine whether to play a note or not. After deciding to play a note, we sample from the replay probability to determine if the note should be re-attacked. Dynamics level is directly used from the model given that the note is played.

We did not use any priming method to generate music, which sometimes resulted in the model generating long initial periods of silence. To overcome this problem, we implemented an adaptive temperature adjustment method to increase the temperature \(T\) of our output sigmoid functions proportional to how many time steps of silence the model produces: \(T = 0.1 t + 1\) where \(t\) is the number of consecutive prior time steps of silent outputs the model has produced. Whenever the model produces a non-silent output at a time step, \(T\) is reset to 1. This prevented the model from outputting long periods of silence.

\section{Evaluation}
\subsection{Quality Analysis}
To evaluate the quality of DeepJ's music generation, we conducted a subjective experiment using an evaluation method similar to the evaluation used in \textit{Sequence Tutor} \cite{rl_tuner}. The goal of this experiment was to evaluate how DeepJ's music generation compared to that of Biaxial architecture's generation. We conducted a user survey, using Amazon Mechanical Turk, in which 20 users were asked to listen to 15 pairs of random samples between outputs produced by DeepJ and Biaxial and to choose the one they preferred more (the chart compares specific styles to random Biaxial pieces).

\begin{figure}
	\centering
    \includegraphics[width=0.95\linewidth]{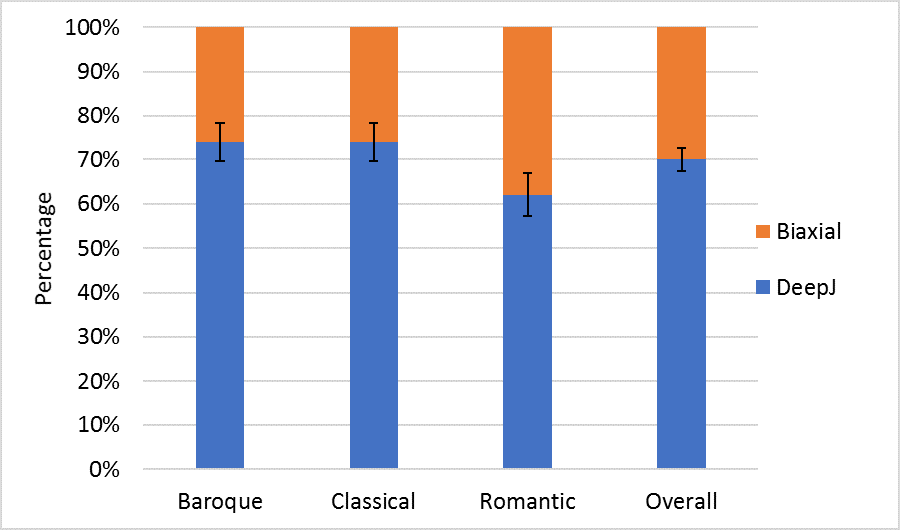}
    \caption{User preferences between DeepJ and Biaxial}
    \label{fig:preference_chart}
\end{figure}

\begin{figure}
  \centering
  \includegraphics[width=0.95\linewidth]{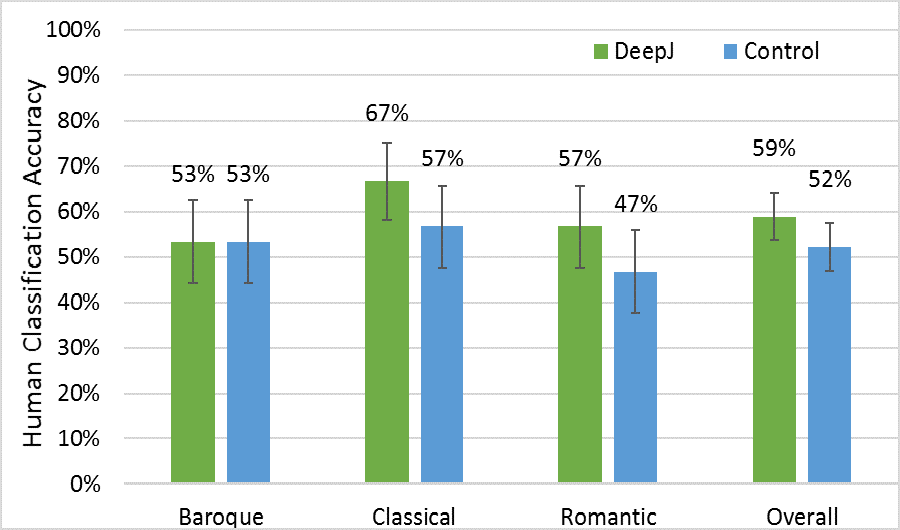}
  \caption{Human accuracy on classifying samples into genres}
  \label{fig:classification_chart}
\end{figure}

Users preferred DeepJ's compositions over Biaxial's 70\% \(\pm\) 5.16\% (95\% confidence interval). Out of 300 user preferences, 210 selected DeepJ, which shows that DeepJ produces better sounding music than Biaxial. This demonstrates that our methods for learning music dynamics and style are effective in improving the quality of generated music. Volume adds a level of emotion to the music and style makes the output more stylistically consistent, preventing it from changing style in the middle of a piece.

\subsection{Style Analysis}
To evaluate the style diversity of DeepJ, we surveyed 20 individuals with musical backgrounds and asked them to classify music generated by DeepJ as baroque, classical or romantic. The goal of the experiment was to evaluate the extent to which DeepJ can create stylistically distinct music by testing if humans can identify the genres. In the study, participants were given 9 music samples generated by DeepJ or by real composers. Ten of our participants were given control samples and ten were given real samples. 

Participants classified the styles of DeepJ outputs correctly 59\% \(\pm\) 13.36\% (95\% confidence interval) of the time (53 out of 90). The control samples were classified correctly 52\% \(\pm\) 13.56\% of the time (47 out of 90). A statistical hypothesis test with significance level of \(0.05\) and \(z = 0.945\) reveals no statistical difference between the accuracy of classification between DeepJ and human composers. An interesting observation is that participants found it challenging to classify control samples. We suspect this may be due to the fact that classical music sometimes share traits of baroque and romantic compositions, making it difficult to distinguish these two. In contrast, when DeepJ produces music it is forced to mix all composers of a particular genre together, bringing out the average characteristics of the genre, which make them more discriminable. Overall, our data demonstrates that DeepJ produces music with style approximately as distinguishable as those composed by humans.

\begin{figure}[ht]
  \centering
  \fbox{\includegraphics[width=0.2\textwidth]{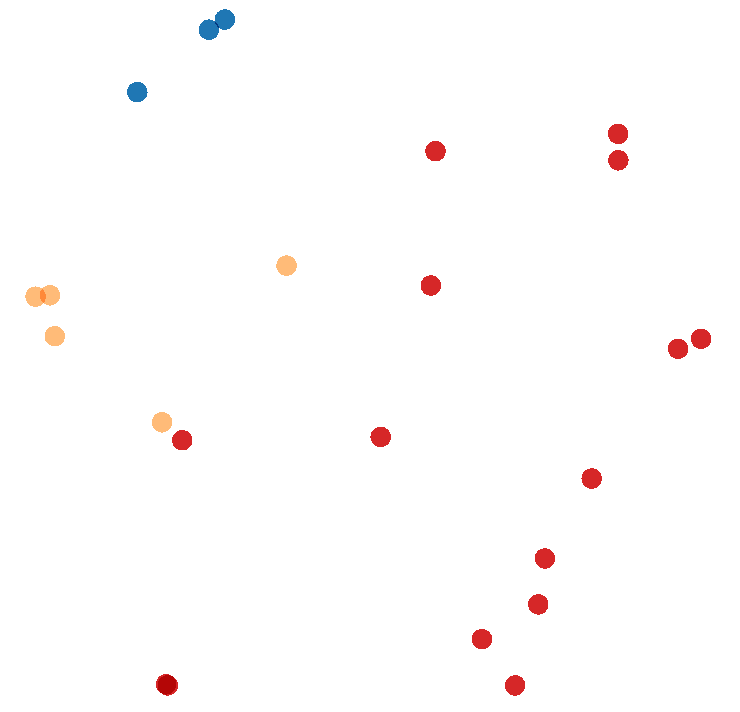}}
  \caption{Style embedding t-SNE visualization with perplexity of 10 and learning rate of 10 after 3000 iterations. Blue, yellow and red dots are baroque, classical and romantic composers respectively.}
  \label{fig:embedding}
\end{figure}

We further analyze the capacity for the model to learn style by visualizing the style embedding space using t-SNE (Figure \ref{fig:embedding}). We notice that composers from similar classical periods tend to cluster together. In the t-SNE visualization (Figure \ref{fig:embedding}) the baroque composers cluster on the top left, classical composer cluster on the left and romantic composers dominate the right. The clustering behavior indicates that DeepJ has learned the similarity and differences between composers. An interesting result to note is that Beethoven, whom we labeled as a classical composer (yellow dot next to the red dot in Figure \ref{fig:embedding}), falls between the cluster centers of classical and romantic composers. This is consistent with the observation that he is known as a composer who represents the transition between the classical and romantic eras.

The generated samples produced by DeepJ also demonstrate that the model has learned artistic style. We discovered that the samples produced under baroque conditioning exhibited counterpoint and polyphonic characteristics similar to Bach's compositions.

\begin{figure}[ht]
  \centering
  \includegraphics[width=0.5\textwidth]{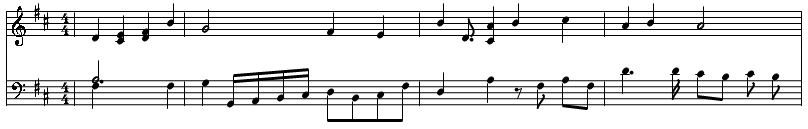}
  \caption{Example baroque output from DeepJ exhibiting fugue characteristics. The second bar introduces a new melody line on the left hand imitating the existing right hand melody.}
  \label{fig:counterpoint}
\end{figure}

Similarly, we also noticed that the output conditioned on classical music demonstrates less complexity and homophony compared to baroque outputs. The output conditioned on romantic period music generally had more freedom in rhythm and contained chromatic harmonies.

An interesting result we found was that the addition of dynamics not only made the output sound more similar to a human playing, it also improved the qualitative output of the model. We believe dynamics provide the model with additional contextual information to make its prediction, such as the emotional qualities the composer attempted to convey through dynamics. Another hypothesis is that dynamics served as a multitask teaching signal that trains the model on a harder task, which leads to better performance \cite{multitask}. Furthermore, we also notice that our model learns the dynamics of the particular styles. Similar to the training data, baroque music generated by DeepJ tends to have more constant dynamics compared to the varying dynamics in classical and romantic outputs.

We encourage the reader to judge the quality and style of our model's samples themselves at \url{https://github.com/calclavia/DeepJ/tree/icsc/archives/v1}.

\section{Conclusion and Future Work}
We introduced a model with the goal of learning musical style using deep neural networks and successfully demonstrated a method of using a distributed representation of style to influence the model to generate music with a given mixture of artist styles. We improved the Biaxial model by adding volume and style, which in turn improved the overall quality of generated music. Our model also solved style consistency problems that were present in the Biaxial architecture.

However, the lack of long term structure and central theme in the generated music is still a problem yet to be solved. It would be interesting to combine reinforcement learning methods from models such as \textit{Sequence Tutor} \cite{rl_tuner} or exploring adversarial methods to train models with better long term structure. In addition, developing a sparse representation of music may be preferred \cite{performance-rnn-2017}, as the representation used by Biaxial is expensive to train.

\bibliographystyle{IEEEtran}
\bibliography{bib}

\end{document}